\documentclass[aps,twocolumn,showpacs]{revtex4}
\usepackage{graphicx}

\begin{document}

\title{\bf Entanglement between two macroscopic fields by coherent
atom-mediated exchange of photons}
\author{C. L. Garrido Alzar, M. Fran\c ca Santos,\cite{byline} and P.
Nussenzveig}
\affiliation{Instituto de F\'\i sica, Universidade de S\~ao Paulo,
Caixa Postal 66318, 05315-970 S\~ao Paulo, SP, Brazil}
\date{\today}

\begin{abstract}
Using two different criteria for continuous variable systems we demonstrated 
that pump and probe beams became quantum correlated in a situation of 
Electromagnetically Induced Transparency in a sample of $^{85}$Rb 
atoms. Our result combines two important 
features for practical implementations in the field of quantum information 
processing. Namely, we proved the existence of entanglement between 
two macroscopic light beams, and this entanglement is intrinsically associated 
to a strong coherence in an atomic medium.
\end{abstract}

\pacs{03.65.Ud, 42.50.-p, 42.50.Lc, 03.67.-a}
\maketitle

In the last two decades, the philosofical and technological impacts of quantum 
correlations or entanglement in multipartite physical systems have been 
accentuated by the theoretical and experimental investigations leading to the 
development of the area of quantum information processing~\cite{bennett}. The 
existence of entanglement, which is an evidence of the non-local character of 
the quantum theory, has been confirmed in quantum optics experiments using 
dichotomics~\cite{aspect} and continuous variable 
systems~\cite{kimble1}. These 
experimental realizations made possible some implementations in the fields of 
quantum information~\cite{kimble,gisin} and computation~\cite{haroche1,ions}. 
Even so, the unavoidable interaction between the physical system used to 
process the quantum information and the environment leads to a loss 
of coherence, and consequently, to a loss of information that introduces 
important limitations to the practicality of the quantum information 
processing technology. Since it is impossible to have isolated 
systems, among the different approaches employed to reduce the influence of 
the environment we find the use of continuous variable systems, specifically, 
intense light beams as in the case of quantum teleportation based on the 
entanglement between twin beams issued from an OPO~\cite{kimble}. In 
that sense, there are propositions, for exemple, for 
quantum cryptography~\cite{ralph}, quantum computation~\cite{lloyd}, dense 
coding~\cite{samuel}, and quantum key distribution~\cite{leuchs}. Another 
approach is to use coherently-prepared atomic media and, in this case, the 
Electromagnetically Induced Transparency (EIT) is a good candidate as it 
has been demonstrated recently by the observation of very slow light pulse 
propagation~\cite{paq} and light storage~\cite{p}. In this context, a 
natural question arises: can we produce entanglement between two 
intense ligth beams in a coherently-prepared atomic medium, combining 
in that way the two mentioned approaches? As we will see through out this 
paper, the answer to this question is yes and this result is very important 
due to the recent huge interest in the applications of 
coherently-prepared atomic media using EIT~\cite{mastko}.

In this paper, we show that three-level atoms can produce
entanglement in two intense travelling light fields, in the
EIT regime. We study this system
from a theoretical point of view, and we show that for some
particular parameters, the two fields present quantum correlations
after interacting with the atoms, even if they are initially
completely uncorrelated. We demonstrate the existence of entanglement 
between these two propagating fields according to two different criteria.

In our model, we consider three-level atoms in a closed $\Lambda$
configuration (ground states $|1\rangle$ and $|2\rangle$, and
excited state $|0\rangle$) interacting with two copropagating fields
treated quantum-mechanically. In the Heisenberg picture, the
electric field operator for the propagating mode $j$ (pumping
laser $j=1$, probe laser $j=2$) is given by the expression
\begin{equation}
 \widehat{\vec{E}}_{j}(t)  =  {\cal E}_{0\omega_{Lj}} \bar{\epsilon}_{j}
 e^{-i \omega_{Lj} t}\hat{A}_{j}(t) + h.c.\ ,
 \label{Eq1}
\end{equation}
where ${\cal E}_{0\omega_{Lj}}$ , $\bar{\epsilon}_{j}$ and
$\omega_{Lj}$ are the amplitude, the polarization direction and
the angular frequency of mode (laser) $j$, respectively.
$\hat{A}_{j}$ ($\hat{A}^{\dag}_{j}$) is the annihilation
(creation) operator and represents the slowly varying amplitude of
the laser field. We take the hamiltonian
\begin{equation}
 \hat{H}_{Lj}=\int_{-\infty}^{+\infty}d\omega\ |G_{j}(\omega)|^{2}
 \hbar\omega\ \hat{a}^{\dag}_{j\omega}\hat{a}_{j\omega}\
 \label{Eq2}
\end{equation}
as the energy source of the interacting field where,
$\hat{a}_{j\omega}$ is the annihilation operator of the field
inside the laser source cavity and, through the non-dimensional
function $G_{j}(\omega)$, we take into account the influence of
the external vacuum modes. This function is determined by the
frequency-dependent reflectivity of the output mirror of the laser
cavity and provides the laser linewidth $\gamma_j$, which we
assume to be constant here, in accordance with the Markov
approximation~\cite{gardiner}. Taking into account the theoretical and 
experimental studies about the laser sources, we took $G_{j}(\omega)$ 
as a lorentzian profile centered at the laser frequency $\omega_{Lj}$, 
allowing the lower integration limit in~(\ref{Eq2}) be taken equal to 
$-\infty$ instead of zero.

The coupling between the source and the propagating mode is given by the 
linear hamiltonian
\begin{equation}
 \hat{H}_{Lj-Cj}=i \hbar\sqrt{\frac{\gamma_{j}}{2\pi}} \int_{-\infty}^{+\infty}
 d\omega\ \hat{a}^{\dag}_{j\omega}\hat{A}_{j}(t)+ h.c.
 \label{Eq3}
\end{equation}
We obtain the interaction hamiltonian for the two light beams
and the atoms, within the usual dipole and rotating-wave
approximations
\begin{eqnarray}
 \hat{H}_{int}=\hbar g_{1}\hat{S}^{+}_{1}(t)\hat{A}_{1}(t) +
 \hbar g_{2}\hat{S}^{+}_{2}(t)\hat{A}_{2}(t) + h.c. \ ,
 \label{Eq4}
\end{eqnarray}
where $g_{1}$ ($g_{2}$) is the atom -- field 1 (field 2) coupling strength,
and $\hat{S}^{+}_{1}$ ($\hat{S}^{+}_{2}$) the slowly varying envelope
of the atomic polarization on the transition
$|1\rangle \leftrightarrow |0\rangle$ ($|2\rangle \leftrightarrow |0\rangle$).

The dynamics of the system is determined by twelve coupled quantum
Langevin equations derived from the Heisenberg equations of
motion. Since we are dealing with macroscopic systems, the quantum
fluctuations of the operators are studied by linearizing them
around their steady-state values and the dynamics of these
fluctuations is described by a matrix linear stochastic
differential equation for the fluctuation operators~\cite{tobe}.
Recently, intensity correlations between the pump and probe fields
in EIT have been measured~\cite{exper}. These can be understood by
inspection of the equations for the fluctuations of one field and
for the corresponding atomic polarization:
\begin{equation}
 \frac{d\delta\hat{A}_{1}(t)}{dt}=-\frac{\gamma_{1}}{2}
 \delta\hat{A}_{1}(t) - i g_{1} \delta\hat{S}^{-}_{1}(t)
 +\sqrt{\gamma_{1}}\delta\hat{A}_{1in}(t)\ ,
 \label{Eq5}
\end{equation}
\begin{eqnarray}
 \frac{d\delta\hat{S}^{-}_{1}(t)}{dt}=-\left(\frac{\Gamma_{1}+\Gamma_{2}}{2}-
 i\ \delta_{L1}\right)\delta\hat{S}^{-}_{1}(t) \nonumber\\
+ i g_{1}w_{1} \delta\hat{A}_{1}(t) +i g_{1}\alpha_{1}\delta\hat{W}_{1}(t)
\nonumber\\
-i g_{2}s^{*}_{12}\delta\hat{A}_{2}(t)
 -i g_{2}\alpha_{2}\delta\hat{S}^{\dag}_{12}(t)
 + \hat{F}_{S1}(t)
 \; .
 \label{Eq6}
\end{eqnarray}


Here we define $\hat{A}_{1in}$ the annihilation operator of the source or 
input field 1, $\Gamma_1$ ($\Gamma_2$) the spontaneous emission rate from 
$|0\rangle \rightarrow |1\rangle$ ($|0\rangle \rightarrow |2\rangle$), 
$\delta_{L1}$ the detuning between field 1 and the corresponding atomic 
transition, $w_1$ the steady-state inversion between states $|0\rangle$ and 
$|1\rangle$, $\alpha_1$ ($\alpha_2$) the steady-state amplitude of field 1 
(field 2), $\hat{W}_1$ the inversion (operator) between states $|0\rangle$ and 
$|1\rangle$, $s^*_{12}$ the steady-state coherence between ground states 
$|1\rangle$ and $|2\rangle$, $\hat{S}^+_{12}$ the coherence operator, 
$\hat{F}_{S1}$ the Langevin fluctuation force. The notation 
$\delta \hat{S}^-_1$ means fluctuations of the corresponding operator. The 
fluctuations of the input $\delta\hat{A}_{1in}$, the interacting 
$\delta\hat{A}_{1}$ and the detected $\delta\hat{A}_{1out}$ fields are related 
by the expression 
$\delta\hat{A}_{1out}=\delta\hat{A}_{1in}-\sqrt{\gamma_{1}}\delta\hat{A}_{1}$.

\begin{figure}[ht]
\includegraphics*[scale=0.42]{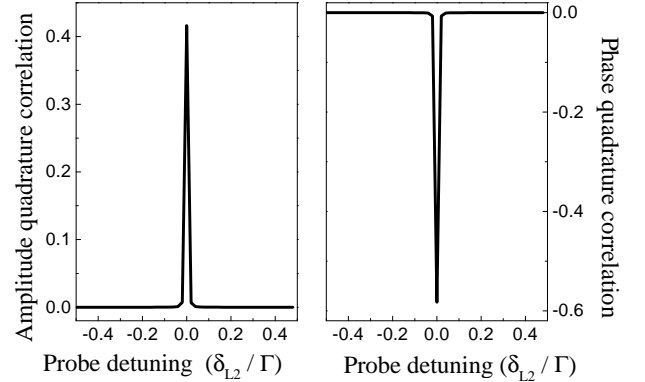}
\caption{Correlation between amplitude and phase quadratures.}
\label{cira2}
\end{figure}

From Eq.~(\ref{Eq6}), we notice that noise correlations between the fields
are created, owing to the coherent effect in the atomic 
medium~\cite{exper}. In Fig.~\ref{cira2} we show the quadrature correlations 
of the fields in the frequency domain as a function of probe detuning for a 
resonant pump. This theoretical prediction (and the following too) corresponds 
to a system of N=$10^{8}$ atoms 
of $^{85}$Rb where the states of the $\Lambda$ configuration are designated as 
follows: $|0 \rangle = |5P_{3/2},F^{'}=3 \rangle$, 
$|1 \rangle = |5S_{1/2},F=3 \rangle$ and 
$|2 \rangle = |5S_{1/2},F=2 \rangle$. The 
pump and probe lasers are taken linearly polarized with equal intensities 
(2.8 mW/cm$^{2}$) and issued from two independent sources with quantum 
fluctuations corresponding to a coherent state. We took the analysis 
frequency $\Omega=\Gamma/6$, where 
$\Gamma=\Gamma_{1}+\Gamma_{2}=2\pi\ 6$ MHz is the total decay rate of the 
rubidium excited state. The correlation, taking values in the interval 
[-1;1], is defined as the ratio between the fields covariance and 
the squared root of the product of the fields' variances. Outside the EIT 
window the fields are completely uncorrelated and, in the EIT condition 
(zero probe detuning), there is the following correlation between the 
fluctuations of the fields quadratures: 
$\delta\hat{Y}_{1out,0} \leftrightarrow \delta\hat{Y}_{2out,0}$ and 
$\delta\hat{Y}_{1out,\pi/2} \leftrightarrow -\delta\hat{Y}_{2out,\pi/2}$. The 
subscript 0 ($\pi/2$) stands for the field amplitude (phase) quadrature and 
the general quadrature fluctuation operator is, for the field 2, given by
$\delta \hat{Y}_{2out,\phi}(t)= \delta \hat{A}_{2out}(t) e^{-i \phi}+ 
\delta \hat{A}^{\dag}_{2out}(t) e^{i \phi}$.

The observed correlations can be interpreted from the propagation dynamics 
of the beams. When the beams have the same intensity, the role ``pump" 
and ``probe" is interchangeable and for a resonant coupling of both 
fields the atomic medium presents exactly the same absorptive and 
dispersive responses for these beams. This regime, that may be called 
electromagnetically mutual induced transparency (EMIT), is broken down 
when we introduce a detuning in one of the fields, creating in this way 
a phase difference between them leading to their decorrelation.

We use two criteria for continuous variable systems to distinguish 
between quantum and classical correlations. We begin our analysis with the 
criterion of the inferred variances, described theoretically in~\cite{reid} 
and experimentally implemented in~\cite{kimble1,kimble}. Let us suppose 
that we are interested in the inferrence of the probe field
amplitude ($\phi=0$) and phase ($\phi=\pi/2$) quadratures from
measurements of the pump field quadratures. In this case, the inferred
variances of the probe quadratures are defined by the equations
\begin{eqnarray}
 \Delta^{2}_{inf}Y_{2,0}(t)\equiv
 \langle\Big(\hat{Y}^{'}_{2out,0}(t)-
 \eta_{0}\ \hat{Y}^{'}_{1out,0}(t)\Big)^{2} \rangle \ ,
 \label{Eq7}\\
 \Delta^{2}_{inf}Y_{2,\pi/2}(t)\equiv
 \langle\Big(\hat{Y}^{'}_{2out,\pi/2}(t)+
 \eta_{\pi/2}\ \hat{Y}^{'}_{1out,\pi/2}(t)\Big)^{2} \rangle \ ,
 \label{Eq8}
\end{eqnarray}
where $\hat{Y}^{'}_{j out,0}(t)=\hat{Y}_{j out,0}(t)-
\langle \hat{Y}_{j out,0}(t) \rangle $ and $\hat{Y}^{'}_{j out,\pi/2}(t)=
\hat{Y}_{j out,\pi/2}(t)-\langle \hat{Y}_{j out,\pi/2}(t) \rangle $ with
$j=1,2$. The parameters $\eta_{0}$ and $\eta_{\pi/2}$ take into account the
non-perfect correlation between the fields and the non-ideal efficiency of 
the measurement procedure. The values of these parameters are taken in order 
to minimize the inferred variances (\ref{Eq7}) and (\ref{Eq8}), allowing the
following criterion, in the frequency domain, for the entanglement of the 
pump and probe fields
\begin{equation}
 [\Delta^{2}_{inf}Y_{2,0}(\Omega)]_{min}
 [\Delta^{2}_{inf}Y_{2,\pi/2}(\Omega)]_{min}<1\ .
 \label{Eq9}
\end{equation}

\begin{figure}[ht]
\includegraphics*[scale=0.42]{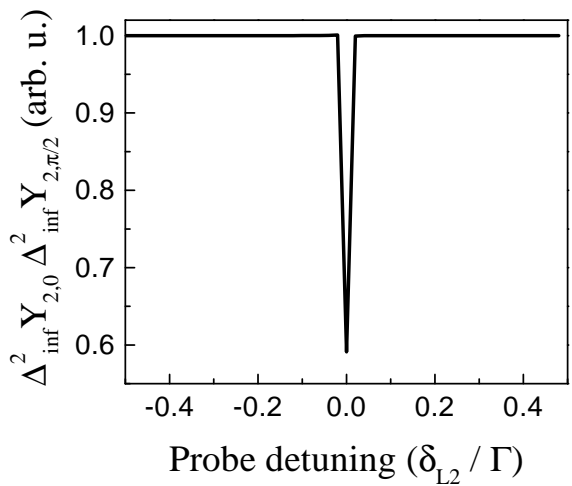}
\caption{Product of the minimal inferred variances of the probe field.}
\label{vari1}
\end{figure}

That is to say, if the product of the inferred variances is less than 1, then
the correlation between the fields has a quantum nature. In Fig.~\ref{vari1}
we show the product of the inferred variances for the probe field. As 
expected, outside de EIT region, the inequality (\ref{Eq9}) is violated since 
the fields are uncorrelated (see Fig.~\ref{cira2}). However, in the EIT 
condition, the pump and probe fields became quantum correlated.

The other criterion used to establish the nature of the fields correlation 
is the theorem of Duan {\it et al.}~\cite{zoller}, which we will abreviate 
as DGCZ. Taking $a=1$, introducing the equivalence between the fields 
quadrature operators and the operators defined in~\cite{zoller} as
$\hat{x}_{1} \Leftrightarrow \hat{Y}_{1 out,0}$,
$\hat{p}_{1} \Leftrightarrow \hat{Y}_{1 out,\pi/2}$,
$\hat{x}_{2} \Leftrightarrow \hat{Y}_{2 out,0}$ and
$\hat{p}_{2} \Leftrightarrow \hat{Y}_{2 out,\pi/2}$, and using the 
commutation relation $[\hat{x}_{j},\hat{p}_{j^{'}}]=2 i \delta_{j j^{'}}$ 
derived from the definition of the quadrature operators, we find the 
following necessary condition to prove that the join state of the pump 
and probe fields is separable
\begin{equation}
 \langle (\Delta\hat{u})^{2} \rangle_{\rho} +
 \langle (\Delta\hat{v})^{2} \rangle_{\rho} \ge 4\ .
 \label{Eq10}
\end{equation}

\begin{figure}[h]
\includegraphics*[scale=0.42]{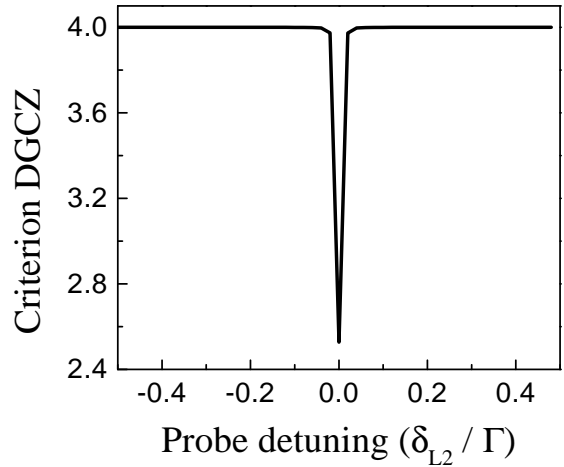}
\caption{Pump-probe entanglement according to the DGCZ criterion.}
 \label{cira4}
\end{figure}

Since this last inequality provides a necessary condition for the 
separability of the join fields state, then its violation is a 
sufficient condition for the inseparability or entanglement 
between the fields. In Fig.~\ref{cira4} we plot the left hand side of 
the inequality (\ref{Eq10}) and, again, outside the EIT window the 
equality is satisfied since the fields are uncorrelated. In the EIT 
condition, the violation of (\ref{Eq10}) indicates that the pump and 
probe fields are entangled. We must point out that both criteria report 
about 40 \% of entanglement of the fields. This amount of quantum 
correlation is limited by the decay rate of the coherence between the 
two ground states.

The predicted quantum correlation between the pump and probe fields in 
the EIT condition is associated to the existence of phase quadrature 
squeezing in both fields. This squeezing, produced in the coherent 
situation, depends on the intensities of the source fields and, as it 
can be observed from the bistability response of the detected fields, 
it is produced at the turning point of the bistability curve and is 
accompanied by an excess noise in the corresponding conjugate quadrature 
amplitude~\cite{tobe}.

These results may be somehow unexpected because it is believed that in 
the EIT condition the field fluctuations are not altered for field 
intensities higher or comparable to the saturation intensity of 
the atomic transition. As we showed theoretically and 
experimentally~\cite{exper} this 
is not the case. In the EIT situation there is a coherent atom-mediated 
exchange of photons between the pump and probe fields that preserves 
their mean intensities and at the same time modifies their  
quantum noise properties creating a correlation between them. This 
modification of the field quantum fluctuations is a direct consequence 
of the strong coherence induced in the atomic medium by the two 
beams. 

So far, the correlation properties of the pump and probe beams
have not been extensively studied in the EIT experiments. Entanglement 
between two single-photon pulses (quantum fields) has been predicted 
before in a coherently-prepared medium by two classical 
beams~\cite{lukin}. In this paper, we show that the \textit{same} 
intense beams used to prepare the transparent nonlinear medium,
in particular circumstances, become entangled even when the 
investigated system is subjet to the influence of a reservoir and 
consequently the quantum correlations are predicted in a system that 
is not ``pure". Since we demonstrated the existence of entanglement 
between the light beams \textit{only}, our result 
suggests that there exists stronger quantum correlations in the 
system \mbox{atom -- pump field -- probe field}. Another remarkable point 
is that the investigation of the correlations and quantum fluctuations 
of the light beams in the EIT situation provides a precise tool to 
determine the natural width of the EIT resonance, and this can be a 
very powerful method to attain the highest sensibility in detecting 
coherent effects in atomic media.

Given the possible technological applications of such intense 
entangled fields, it is our understanding that the statistical 
properties of these fields deserve further experimental investigations 
in the near future. Not only do they open the possibility to use such 
systems as a macroscopic resource for different quantum 
technologies, but they also help understanding the nature of 
entanglement and how it may arise from non-linear couplings in 
macroscopic media. 

Finally, in Fig.~4 we present an experimental setup that can be employed 
to measure the entanglement between the pump and probe fields. Considering 
the light beams have orthogonal polarizations, they can be 
separated using polarizing cube beam splitters and then the variance of 
their quadratures and the correlation between them can be determined 
utilizing homodyne detectors. From practical considerations, the analysis 
frequency must be chosen as low as possible since in this case we have 
more sensibility to detect correlated photons.

\begin{figure}
\includegraphics*[scale=0.38]{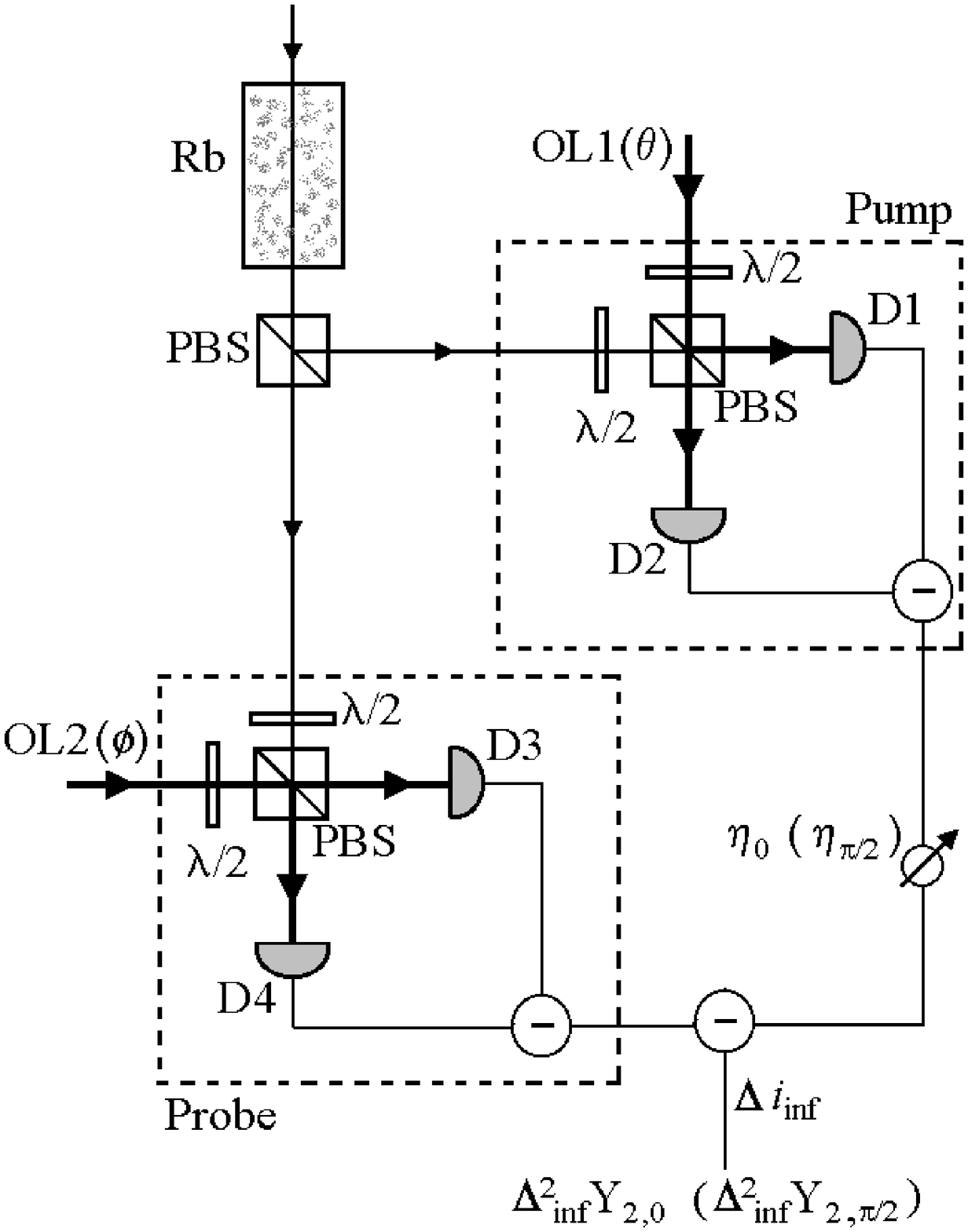}
\caption{Experimental setup to measure the probe field inferred 
 variances. PBS: polarizing cube beam splitter; $\lambda/2$: half-wave plate; 
 D1, D2, D3, and D4: photodetectors; OL1($\theta$) and OL2($\phi$): local 
 oscillators for the pump and probe field, respectively; 
 $\eta_{0}(\eta_{\pi/2})$: controlled-gain amplifier for the amplitude 
 (phase) quadrature.}
\label{varinfexpe}
\end{figure}

M.F.S. would like to thank Prof. S. Salinas for his hospitality at the 
University of S\~ao Paulo. The authors acknowledge the financial support of 
the Brazilian agencies CAPES, FAPESP and CNPq.


\end{document}